\documentclass[twocolumn]{aastex631}
\usepackage{threeparttable} 
\usepackage{graphicx}

\begin{document}

\title{Searching for dual/lensed quasar candidates with spectroscopic surveys}

\author[0000-0002-0539-8244]{Xiang Ji$^\star$}
\affil{Shanghai Astronomical Observatory, Chinese Academy of Sciences, 80 Nandan Road, Shanghai 200030, China}
\author{Zhen-Ya Zheng$^\star$}
\affil{Shanghai Astronomical Observatory, Chinese Academy of Sciences, 80 Nandan Road, Shanghai 200030, China}
\affil{School of Astronomy and Space Sciences, University of Chinese Academy of Sciences, No. 19A Yuquan Road, Beijing 100049, China}

\correspondingauthor{Xiang Ji; Zhen-Ya Zheng}
\email{$^\star$: jixiang@shao.ac.cn; Zhengzy@shao.ac.cn}

\author{Ru-Qiu Lin}
\affil{Shanghai Astronomical Observatory, Chinese Academy of Sciences, 80 Nandan Road, Shanghai 200030, China}
\affil{School of Astronomy and Space Sciences, University of Chinese Academy of Sciences, No. 19A Yuquan Road, Beijing 100049, China}

\author{Hai-Cheng Feng}
\affil{Yunnan Observatories, Chinese Academy of Sciences, Kunming 650216, Yunnan, China}

\begin{abstract}

Dual and lensed quasars are valuable astrophysical targets in many aspects. Dual quasars, considered as the precursors of supermassive black hole binaries, can provide crucial insights into how black hole mergers drive the growth of supermassive black holes and influence the evolution of galaxies. Lensed quasars, formed by the gravitational deflection of a background quasar's light by a massive foreground object, can address key cosmological questions, particularly in refining measurements of the Hubble constant. Despite their significance, the number of confirmed dual and lensed quasars remains limited. Here in this work, we propose a systematic search for dual/lensed quasars using broad emission line profile diagnostics. Our parent sample consists of spectroscopic quasars from the Dark Energy Spectroscopic Instrument Early Data Release (DESI EDR) and the SDSS DR17 catalog. We identify 30 lensed quasar candidates with similar broad emission line profiles, as well as 36 dual quasar candidates with different profiles. Cross-matching these 66 targets with the HST archival database, we find four overlapping targets, including three previously reported lensed quasars and one newly identified dual quasar candidate. We estimate the black hole masses for the two cores in the same system. The mass ratios are similar in the lensed quasar scenario but vary widely for dual quasars, consistent with the physical nature of these two types. In particular, we identified a dual quasar candidate with the mass ratio exceeding 100 times. We aim to discover more dual/lensed quasar candidates using our method with the upcoming future spectroscopic surveys. 

\end{abstract}


\section{Introduction} \label{sec:intro}

It is widely recognized that most massive galaxies host supermassive black holes (SMBHs) at their centers. According to modern models of galaxy formation and evolution \citep[e.g.,][]{Begelman1980, Yu2002}, supermassive black hole binaries (SMBHBs) are expected to be prevalent in the centers of many massive galaxies, especially during galaxy mergers \citep[e.g.,][]{Volonteri2003, Chen2020}. Studying SMBHBs across a range of separations is essential to understand the galaxies' merging processes, the growth and evolution of SMBHBs, and the origins of gravitational wave radiation \citep[e.g.,][]{Sesana2009, Chen2020}.

At kiloparsec scales, a dual quasar scenario arises when two merging galaxies each trigger active galactic nucleus activity in the center. Significant efforts have been made to search for dual quasars on kiloparsec scales \citep[e.g.,][]{Liu2010a, Comerford2009, Shen2011, Ge2012, Koss2012, Comerford2015, Liu2018} over the past several decades. Numerous potential dual quasar candidates have been identified based on the double-peaked features of [O III] emission lines \citep[e.g.,][]{Liu2010b, Comerford2012}, although only a small fraction of these have been confirmed to be genuine dual quasars by follow-up spectroscopic observations \citep[e.g.,][]{Shen2011, Comerford2015}.

Another category of quasar pairs, known as lensed quasars, can mimic dual quasars in appearance. However, unlike genuine dual quasars, lensed quasars are multiple images of a background quasar caused by gravitational lensing due to a massive object, such as a galaxy or cluster, along the line of sight \citep[e.g.,][]{Schneider1992}. Lensed quasars are valuable tools for studying various properties of both the quasars and the foreground lensing galaxies \citep[e.g.,][]{Ding2017, Stacey2018}, such as probing the geometry and kinematics of quasar accretion disks \citep[e.g.,][]{Blackburne2011, Blackburne2015}, exploring the broad-line regions of quasars \citep[e.g.,][]{Sluse2011}, investigating the properties of dark matter halos \citep[e.g.,][]{Oguri2004}, and even constraining the Hubble constant \citep[e.g.,][]{Suyu2017, Bonvin2017}.

Recent research has shown that Gaia's high-precision astrometric data can be effectively utilized to search for both dual and lensed quasars. Several methods have been proposed, including 'multiplicity' \citep[e.g.,][]{Ji2023}, 'varstrometry' \citep[e.g.,][]{Shen2021, Chen2022}, and Gaia Multi-Peak (GMP) \citep[e.g.,][]{Mannucci2022}. Even with these methodological improvements, the confirmed number of dual and lensed quasars remains scarce, with only a few hundred dual and lensed quasars identified so far. This scarcity continues to pose challenges to our understanding of the formation and evolution of SMBHB.

However, the situation is expected to improve with the ongoing data releases from spectroscopic surveys, particularly from the Dark Energy Spectroscopic Instrument (DESI) \citep[e.g.,][]{DESI2022}. DESI is designed to explore the nature of dark energy through spectroscopic measurements of millions of stars, galaxies, and quasars. The Early Data Release (EDR) of DESI comprises spectra from nearly 1.8 million distinct targets, approximately 95,000 of which are quasars. Future releases are expected to include spectra for around 3 million quasars \citep[][]{Chaussidon2023}, significantly enhancing resources for identifying dual or lensed quasars.

In this work, we report newly identified dual and lensed quasar candidates selected from DESI EDR and SDSS DR17 data with a systematic search approach. The structure of this paper is as follows. In Section \ref{sec:data}, we introduce the data sets and the target selection strategy. In Section \ref{sec:methods}, we describe the systematic search process, which includes spectral fitting procedures, target classification strategy, and the method for estimating black hole masses. In Section \ref{sec:results}, we present the sample of selected dual and lensed quasar candidates. In Section \ref{sec:discussions}, we present comparisons of the spectra from two cores in a previously reported lensed quasar and a newly selected dual quasar candidate as examples. We further discuss our selection strategy and review available HST images to enhance classification accuracy, while comparing black hole mass ratios within the same system. Finally, we summarize the main conclusions in Section \ref{sec:conclusion}.

\section {Data and sample} 
\label{sec:data}

DESI commenced its inaugural five-year mission in December 2020. Throughout this timeframe, DESI aims to acquire spectra from stars, galaxies, and quasars of over 14000 $deg^{2}$. The targets for DESI are categorized into five primary classes \citep[e.g.][]{DESI2022}, including the Bright Galaxy Survey (BGS) targets, luminous red galaxies(LRGs), emission-line galaxies(ELGs), quasars(quasi-stellar objects, QSOs), and stars in the Milky Way Survey(MWS). In this paper, we predominantly address the initial public release of the DESI spectroscopic dataset, known as the DESI EDR \citep{DESI2023}. The data in this release originate from the "Survey Validation"(SV) of DESI that preceded the main survey. The SV primarily serves to verify target selection algorithms, assess operational capabilities, and validate survey operation procedures and the final target selection. 

Our parent quasar sample comprises both spectroscopically confirmed quasars from SDSS DR17 and those from DESI EDR. The selection process consists of two main steps:

\begin{itemize}
    \item Considering that the typical fiber sizes for DESI and SDSS are 1.5 \arcsec, distinguishing between two targets closer than 1.5 \arcsec is challenging. Thus, we select targets for which the two detections of DESI or SDSS are within the radius of 1.5 \arcsec to 5 \arcsec. Due to the fact that the typical fiber sizes of DESI and SDSS are 1.5 \arcsec, we consider that it is hard to separate two targets less than 1.5 \arcsec. 

    \item We require that there are two distinct cores present in the DESI images.

\end{itemize}

\begin{figure}[ht]
\centering
\includegraphics[scale=0.7]{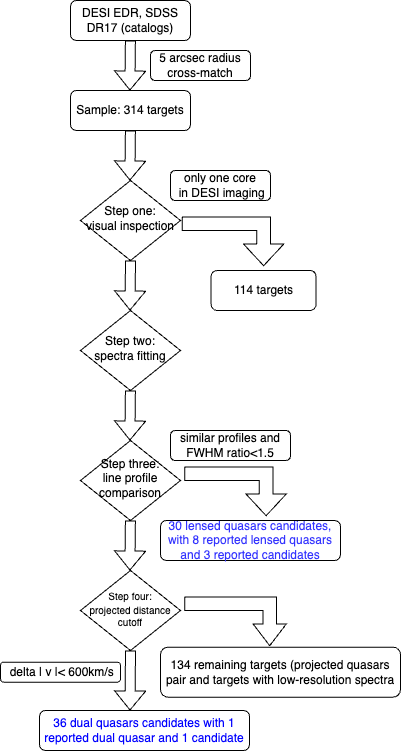}
\caption{
Illustration of the process for selecting and categorizing targets within our sample, with each phase indicating the quantity of chosen targets.
\label{fig:f1}}
\end{figure}

During the initial selection phase, we identified 314 potential targets. A visual inspection led to the exclusion of 114 candidates because only one core was visible in DESI images, while the other was either too faint or missing. This left us with 200 targets for further classification. Furthermore, we recognize 13 targets as previously documented lensed or dual quasar candidates from earlier works, detailed in Table \ref{table:tb1}, which will act as benchmarks in future classification processes. The schematic of our final selection procedures is shown in Figure \ref{fig:f1}.

\begin{table*}[]
\centering
\caption{Basic properties of the documented lensed/dual quasars (candidates).}
\begin{tabular}{lllllll}
\hline \hline
Source id                & RA               & DEC              & Redshift     & Separation(")   &Class              & Reference  \\ \hline
SDSS J121646.05+352941.5 & 184.19188        & 35.494863        & 2.017 & 1.555 & lensing               & [1]  \\
                         & 184.19135        & 35.494881        & 2.006 &       &                       &      \\
SDSS J125418.94+223536.5 & 193.57896        & 22.593497        & 3.649 & 1.562 & lensing               & [1]  \\
                         & 193.57939        & 22.593322        & 3.650  &       &                       &      \\
SDSS J133907.13+131039.6 & 204.77974        & 13.177676        & 2.239 & 1.692 & lensing               & [2]  \\
                         & 204.78014        & 13.177413        & 2.236 &       &                       &      \\
SDSS J140012.77+313454.1 & 210.05322        & 31.581705        & 3.314 & 1.719 & lensing               & [2]  \\
                         & 210.05355        & 31.581319        & 3.316 &       &                       &      \\
SDSS J111816.95+074558.1 & 169.57063        & 7.7661636        & 1.733 & 2.266 & lensing               & [3]  \\
                         & 169.57024        & 7.7666605        & 1.732 &       &                       &      \\
SDSS J100128.61+502756.9 & 150.36921        & 50.465805        & 1.841 & 2.898 & lensing               & [4]  \\
                         & 150.36813        & 50.466224        & 1.846 &       &                       &      \\
SDSS J120629.64+433217.5 & 181.62353        & 43.538219        & 1.791 & 3.003 & lensing               & [4]  \\
                         & 181.62355        & 43.539053        & 1.794 &       &                       &      \\
SDSS J100434.91+411242.8 & 151.14549        & 41.211895        & 1.738 & 3.765 & lensing               & [5]  \\
                         & 151.14503        & 41.210908        & 1.731 &       &                       &      \\
SDSS J081254.82+334950.2 & 123.22843        & 33.830613        & 1.500   & 1.833 & lensing candidate     & [6]  \\
                         & 123.22845        & 33.831122        & 1.500   &       &                       &      \\
SDSS J074013.44+292648.3 & 115.05602        & 29.446782        & 0.978 & 2.661 & lensing candidate     & [6] \\
                         & 115.05594        & 29.446046        & 0.977 &       &                       &      \\
SDSS J221208.05+314416.2 & 333.03380         & 31.73856        & 1.709 & 2.684 & lensing candidate     & [7] \\
                         & 333.03356        & 31.73785        & 1.714 &       &                       &      \\
SDSS J171322.589+325628.02 &    258.34412	& 32.941116        & 0.102 & 4.249 & dual AGN              & [8] \\
                         & 258.34552	       & 32.94123         & 0.101 &       &         
              &      \\
SDSS J121405.12+010205.1 & 183.52129 & 1.03533 & 0.494 & 2.179 & dual quasar candidate & [9] \\
                         & 183.52138 & 1.03473 & 0.492 &       &                       &     \\
\hline \hline
\end{tabular}
\begin{tablenotes}
% \footnotesize
\item
References: 
[1] \cite{Rusu2016}; [2] \cite{Inada2009}; [3]\cite{Weymann1980}; [4]\cite{Oguri2005};
[5]\cite{Inada2005}; [6]\cite{Lemon2018};  [7]\cite{Lemon2019};  [8] \cite{Zhang2021};
[9] \cite{Silverman2020};

\end{tablenotes}
\label{table:tb1}
\end{table*}

\section{Methods}
\label{sec:methods}

In this section, we outline our approach to classifying and analyzing the properties of selected targets. In Section \ref{sec:spectra_fitting}, we describe the procedures for processing the spectra using the Python-based software PyQSOFit, particularly focusing on the comparison of the broad line profile shapes between different cores within the same system. In Section \ref{sec:classification}, we delineate our classification method derived from the results of the spectral fitting. Finally, we introduce the empirical relationship used to estimate the black hole masses of the two cores within the same system in Section \ref{sec:BH_mass}.

\subsection{Spectra fitting}
\label{sec:spectra_fitting}

Quasars that are gravitationally lensed produce multiple images as the light from a distant quasar is deflected because of the presence of a galaxy or cluster positioned close to the line of sight. In such systems, the flux ratio between the spectra of these multiple cores tends to be relatively constant as a function of wavelength or slightly changed because of varying dust extinction or microlensing effects in the different light paths of these images. Distinguishing dual quasars from lensed ones often involves comparing the flux ratio between multiple cores in the same system, as detailed in previous studies \citep[e.g.,][]{Lemon2019, Lemon2023}. However, differentiating these types can be problematic due to possible optical variability across different observation times, complicating the identification of the source of differences between cores within the same system.  To address this issue, we carry out additional comparisons using spectra fitting results.

We employ PyQSOFit \citep[e.g.,][]{Guo2018} for the analysis of quasar spectra. This software processes input data, including observed-frame wavelength, flux density, error arrays, and redshift information. The fitting is performed in the rest frame, producing optimal fitting parameters and user-specified quality-check plots. The software package encompasses a main routine, Fe II templates, a list of line fitting parameters, host galaxy templates, and a dust reddening map to extract spectral measurements from raw FITS files. The number of Gaussian components for each particular broad emission line can be manually adjusted to yield the best-fitting results. The main fitting procedures are described in \citet{Shen2011}. Moreover, PyQSOFit is capable of conducting Monte Carlo or MCMC estimation of the measurement uncertainties of the fitting results. The Full Width at Half Maximum (FWHM) and its uncertainty of each specific line are derived accordingly. These comprehensive fitting results enable comparison of emission line profiles, evaluation of peak flux ratios, and comparison of the FWHM of the same broad emission line across the two cores within the same system.

\subsection{Target Classification}
\label{sec:classification}

In general, lensed quasars display comparable profiles in their broad emission lines. 
The continuum flux ratio among various lensed images can be influenced by microlensing effects \citep[e.g.,][]{Inada2006} or absorption features of lensing galaxies \citep[e.g.,][]{Ellison2004}. For our sample, the spectra for the two cores within the same system were not observed simultaneously. As a result, the slope of the continua can vary due to the intrinsic variation of the QSO itself. Nevertheless, the emission line profiles and the FWHM should exhibit similar features in the lensed quasar scenario. We derive FWHM selection criteria based on the distribution of previously reported lensed quasars.

As suggested by \citet{Chen2022}, the velocity difference between the two cores in dual quasars is typically less than 2000 km/s. However, this criterion may be too broad, especially for high-redshift quasars. In this work, we enforce a more stringent constraint, restricting the intrinsic velocity difference to be less than 600 km/s in the quasar rest frame, to better distinguish dual quasars from projected pairs. Based on the refined analysis, we classify our sample into the following three categories:

\begin{itemize}
\item Lensed quasar candidates: Systems are categorized as lensed quasar candidates when the line profile shapes and the FWHM of the emission lines are similar;

\item Dual quasar candidates: If the line profiles or the FWHM of the emission lines are different, with the intrinsic velocity offset smaller than 600km/s, those systems are classified as dual quasar candidates.

\item Projected quasar pairs: For cases where the line profiles and the FWHM of the emission lines are different, with velocity offsets larger than 600km/s. Those systems are identified as projected quasar pairs.

\end{itemize}

For targets that have broad emission lines with low signal-to-noise ratios, leading to large uncertainties in the FWHM measurements, we temporarily exclude such cases from the analysis, as further data collection is necessary for accurate results.

\subsection{Black hole mass estimating}
\label{sec:BH_mass}

Regarding the central black hole mass measurement for AGNs and quasars, the most commonly used approach is the reverberation mapping of broad emission lines \citep[e.g.,][]{Vestergaard2006,Vestergaard2009}. This technique offers two key advantages: (1) it is independent of angular resolution, making it suitable for distant objects, and (2) it establishes straightforward empirical relationships that serve as reliable secondary indicators, allowing for the estimation of black hole masses for a large sample of AGNs and quasars from single-spectroscopic observations.

Generally, the relationship between the central black hole mass and a specific broad emission line can be expressed as:

\begin{equation}
log M_{BH} = A + Blog\lambda L_{\lambda} + 2log(FWHM)
\end{equation}

where $M_{BH}$ is expressed in $M_{\sun}$, $\lambda L_{\lambda}$ is the unit of $10^{44}$ erg/s, and FWHM is given in 1000 km/s. For the line $H_{\beta}$, the coefficients A and B take the values 6.91 and 0.5, respectively, with the continuum at 5100 Å \citep{Vestergaard2009}. In the case of MgII, A and B are 6.86 and 0.47, and the continuum is measured at 3000 Å \citep{Vestergaard2006}. For the CIV line, the corresponding values are 6.66 and 0.53, with the continuum measured at 1350 Å \citep{Vestergaard2006}. Using these equations, we can calculate the black hole mass for the different cores in a the same system, accounting for their respective redshifts.

\section{Results} 
\label{sec:results}

% \subsection{Selected dual/lensed quasar candidates}
% \label{sec:result_class}

In this section, we introduce the sample of potential dual and lensed quasars chosen using the methods detailed in Section \ref{sec:methods}. The specifics of our selection criteria are elaborated in Section \ref{sec:discussions}. Following the assessment of spectral data and FWHM comparisons, we identified 36 potential dual quasars, including 1 confirmed dual quasar and 1 known dual quasar candidate, along with 30 potential lensed quasars, which comprise 8 confirmed lensed quasars and 3 known lensed quasar candidates. A summary of the chosen lensed and dual quasar candidates is presented in Table \ref{table:tb2} and Table \ref{table:tb3}, respectively.

\begin{figure*}
\plotone{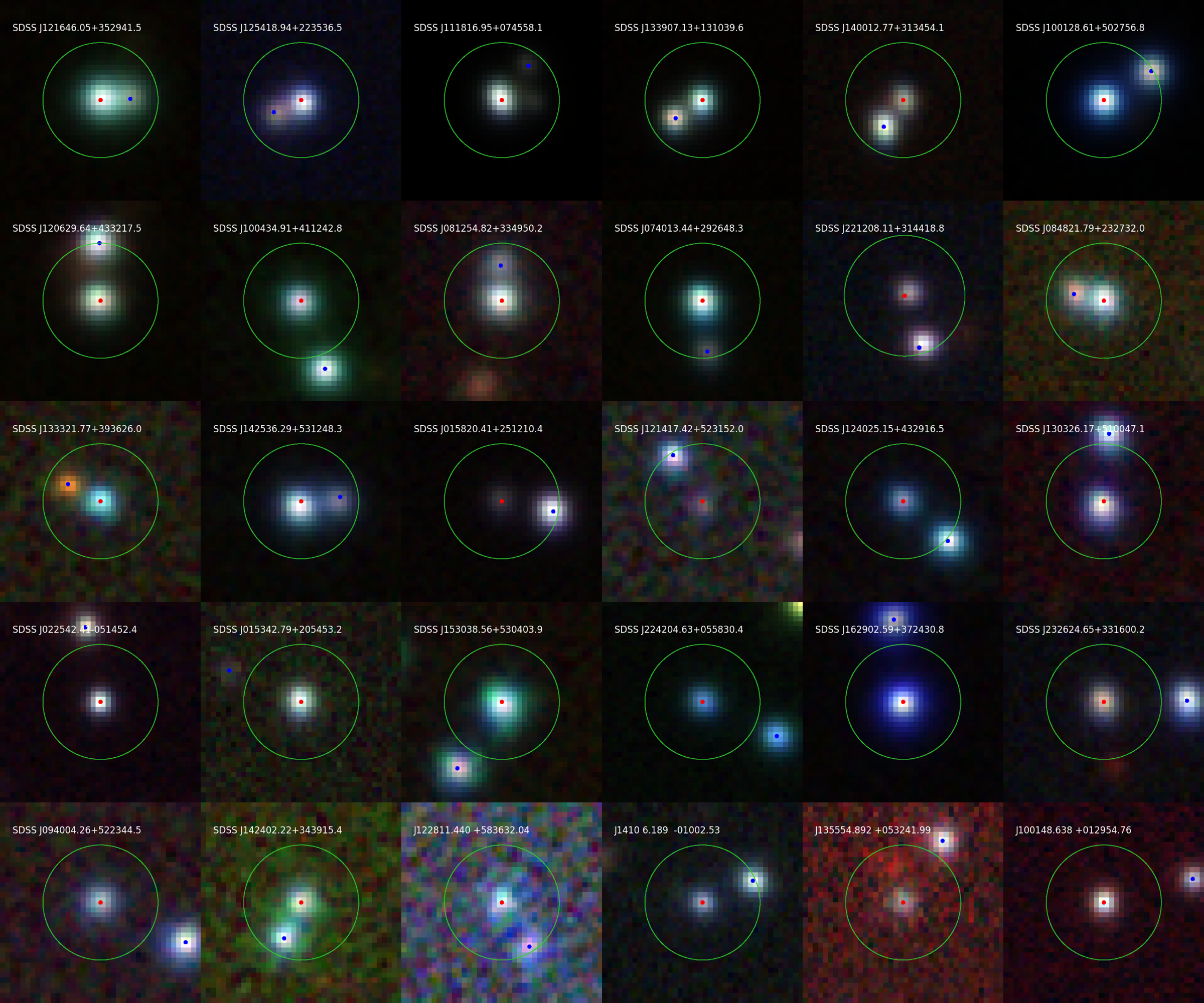}
\caption{
DESI composite color images (DESI i band in red, DESI r band in green, and DESI g band in blue) for the selected dual quasar candidates. For each image, the size is 10\arcsec\  $\times$ 10 \arcsec\ , with the source IDs indicated at the upper left corner. The center of the green circle in each panel is from the SDSS/DESI coordinates, with a radius of 3 \arcsec. In addition, we also mark the detections from \textit{DESI} in red dots in each panel. For the reported lensed quasar candidate SDSS J221208.11+314418.8, there's no available image from DESI, we use the Pan-STARRs g,r and i bands instead.} 
\label{fig:f2}
\end{figure*}

\begin{figure*}
\plotone{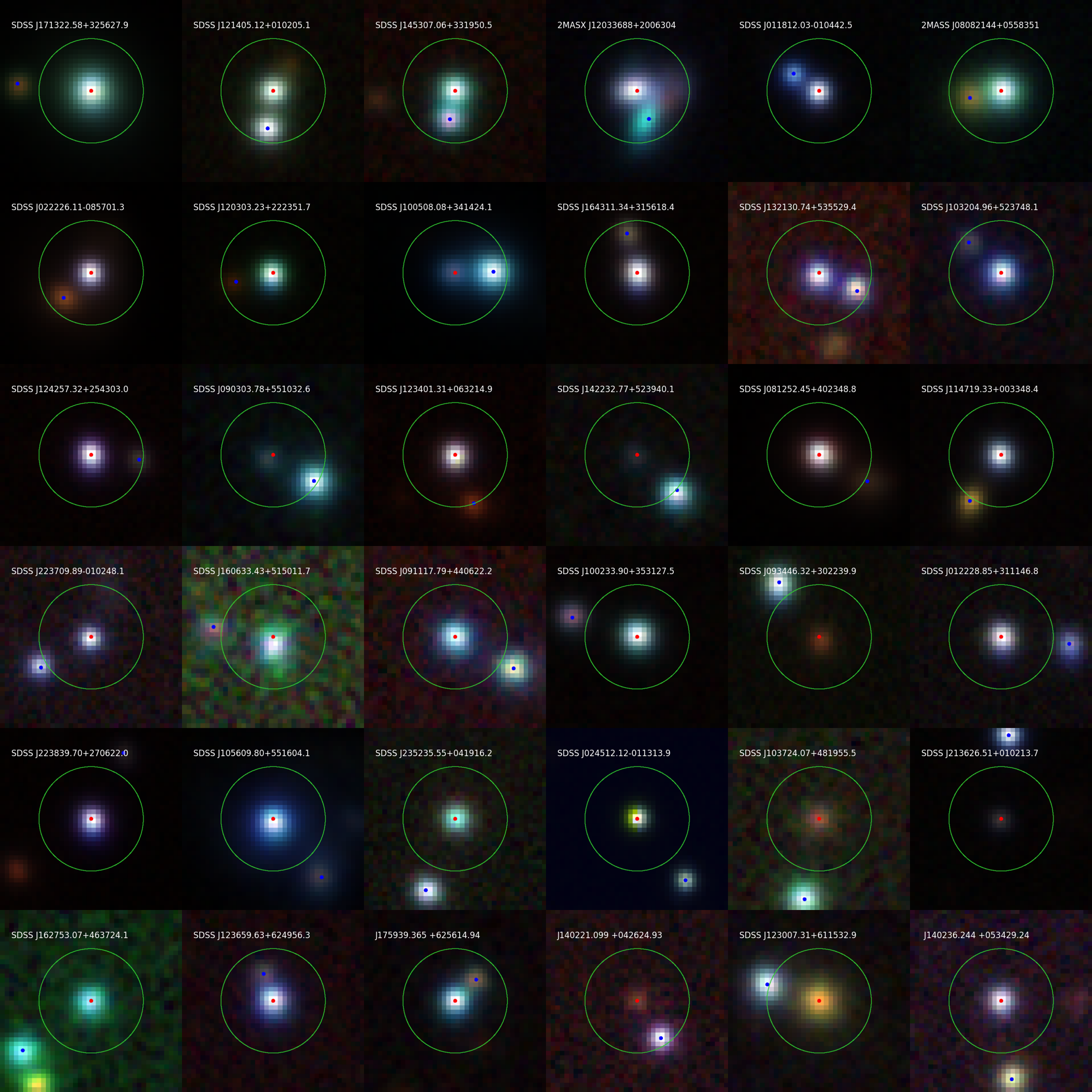}
\caption{
Legends similar to that in Figure \ref{fig:f2} but for the dual quasar candidates from Table \ref{table:tb3}.
\label{fig:f3}}
\end{figure*}

Figure \ref{fig:f2} displays the DESI optical composite color images of the chosen lensed quasar candidates. The spectroscopically identified cores from DESI or SDSS are highlighted with blue and red dots for reference. It is observed that even in reported lensed quasars, the images fail to show the foreground lensing galaxies. Future image decomposition results are essential to better uncover these foreground lensing galaxies. One the other aspect, some of the cores within the same system exhibit similar colors, while others do not. This may be attributed to the varying absorption characteristics along different light paths. Figure \ref{fig:f3} is similar to Figure \ref{fig:f2} but for the selected dual quasar candidates. Although all the targets in both Figure \ref{fig:f2} and Figure \ref{fig:f3} show two distinct cores from DESI imaging, their nature is still elusive, and further high spatial-resolved spectroscopic confirmation is needed to finally pin them down from alternatives.

\section{discussions} 
\label{sec:discussions}

In this work, we propose a systematic approach to search for dual and lensed quasar candidates by utilizing spectroscopic data obtained from DESI EDR and SDSS DR17. Our classification method mainly depends on the examination of the characteristics of the emission lines and the comparison of the FWHM of broad emission lines. 

\subsection{Efficiency of our method}

In this section, we show the example of dual/lensed quasars in section \ref{sec:result_ex} and display the spectral fitting results in section \ref{sec:result_fit}. We then compare the FWHM of the broad emission lines, which serves as the basis for further classification of the remaining targets in our sample in Section \ref{sec:result_fwhm}. We provide a comprehensive discussion of our method in Section \ref{sec:result_over}.

\subsubsection{Example of the dual/lensed quasars}
\label{sec:result_ex}

In general, long-slit spectroscopy is an efficient method for distinguishing dual quasars from lensed quasars by analyzing flux ratios and broad emission line profiles of the two cores within a system. For accurate results, simultaneous spectral observations of both cores are necessary to avoid uncertainties from seeing conditions or target variability over different times. In our sample, although both cores in the same system are spectroscopically confirmed quasars, the different observational times make it challenging to determine whether a system is a lensed quasar or a dual quasar based solely on SDSS/DESI spectra and re-binned flux ratio comparisons.

\begin{figure*}[ht]
\plotone{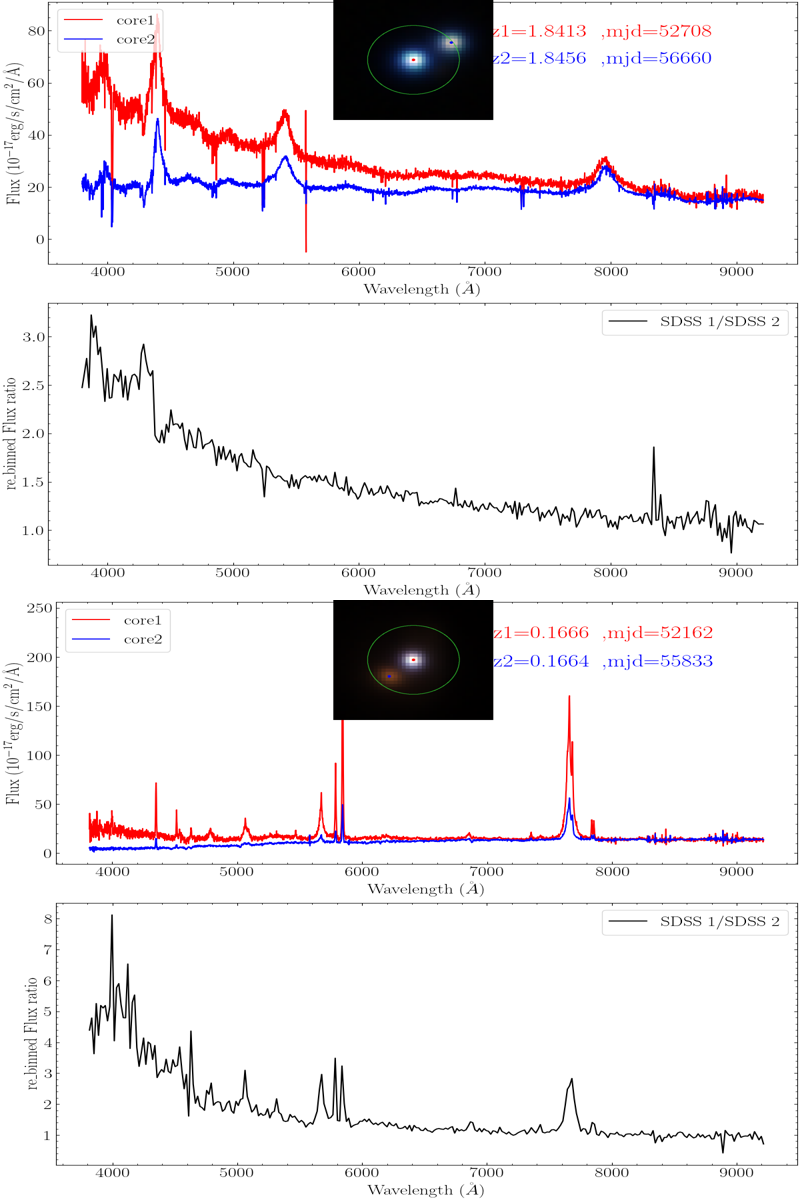}
\caption{
Spectra comparison for the reported lensed quasar SDSS J1001 +5027 (top two panels) and the newly selected dual quasar candidate SDSS J0222 -0857 (bottom two panels). In the first and third rows, the red and blue lines represent for the spectra of the cores marked in red and blue dots of the DESI imaging in the inner panel, respectively. The green circle in the inner panel represents a radius of 3 \arcsec for reference. The second and fourth rows display the re-binned flux ratio between the two cores in the respective system.
\label{fig:f4}}
\end{figure*}

Figure \ref{fig:f4} presents the complete SDSS spectra and the comparisons of re-binned flux ratios for the identified lensed quasar SDSS J1001+5027 and a potential dual quasar SDSS J0222-0857, shown in the top two and bottom two panels, respectively. For the lensed quasar SDSS J1001+5027, the re-binned flux ratio shows non-uniformity throughout the spectrum. This discrepancy could be attributed to microlensing effects or absorption features from the lensing galaxy. Moreover, since the two SDSS spectra were not observed simultaneously, variations in the source quasar's magnitude between different observation times might contribute to differences in the continua. In the case of the dual quasar candidate SDSS J0222-0857, distinct differences are observed in the broad emission line profiles of $H_{\delta}$ and $H_{\gamma}$ between the two cores. The situation is different from the lensed quasar scenario.

\subsubsection{Spectral fitting results}
\label{sec:result_fit}

\begin{figure*}
\plotone{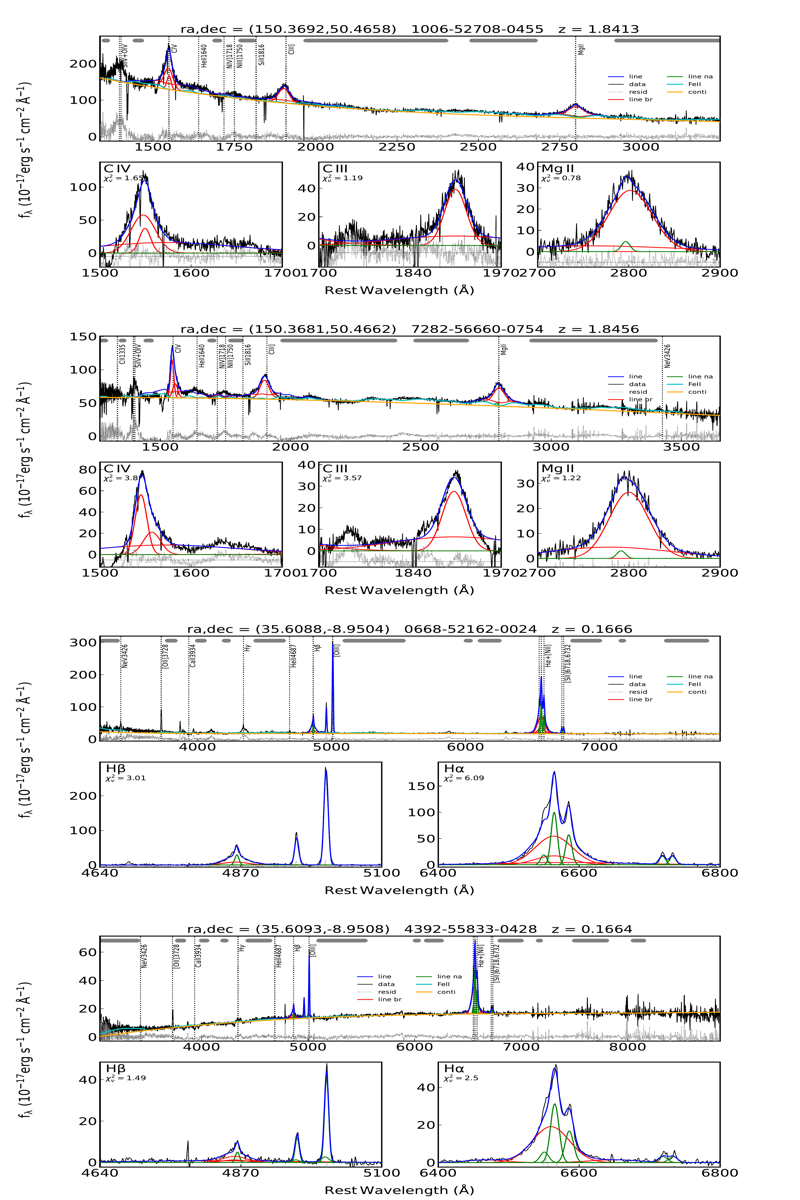}
\caption{
Spectra fitting results of the two cores for the lensed quasar SDSS J1001 +5027 (top two panels) and the newly selected dual quasar candidate SDSS J0222 -0857 (bottom two panels), respectively. In the first two rows, the top panel corresponds to the red core and the bottom to the blue core, as depicted in the inner section of the first row in Figure \ref{fig:f4}. In the lower two rows, the upper panel corresponds to the red core and the lower to the blue core, as indicated in the inner portion of the third row in Figure \ref{fig:f4}. For each core, the fitting details include the continuum and specific emission lines such as CIV, MgII, $H_{\alpha}$, and $H_{\beta}$, with respect to the object's redshift.
\label{fig:f5}}
\end{figure*}

Figure \ref{fig:f5} illustrates the spectra fitting results for the two targets shown in Figures \ref{fig:f4}. Specifically, the first two rows correspond to the two cores associated with the lensed quasar SDSS J1001 +5027. For each core, we provide the fitting results regarding both the continuum and the associated broad emission line profiles. The FWHM and its uncertainties are calculated based on the fitting results as explained in Section \ref{sec:spectra_fitting}. Utilizing the spectral fitting methods outlined in Section \ref{sec:methods}, we applied these techniques to fit all targets in our study.

\begin{figure*}
\plotone{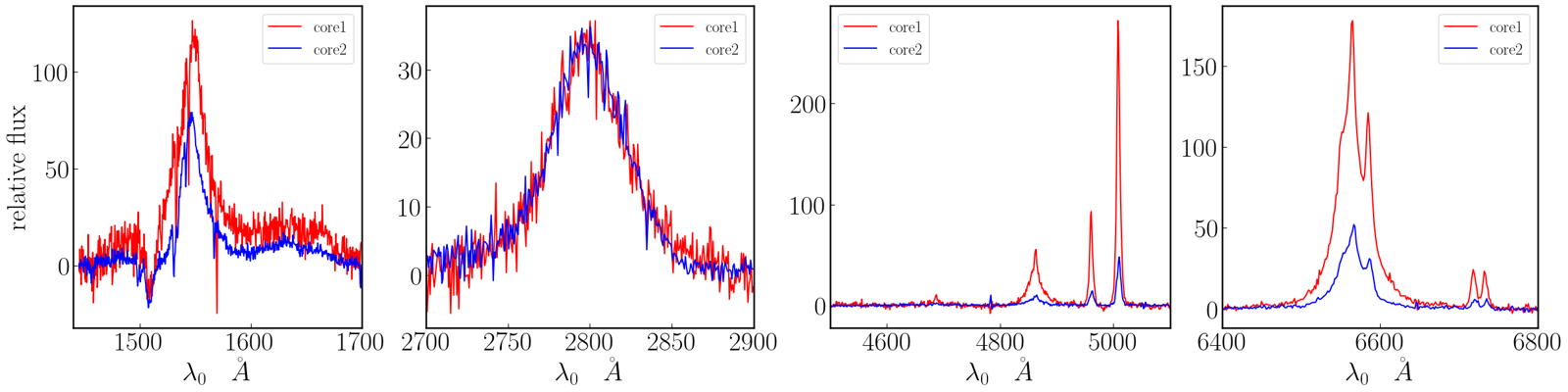}
\caption{
FWHM comparison results for the broad emission line profiles of CIV and MgII for the lensed quasar SDSS J1001+5027 (left two panels) and $H_{\beta}$ and $H_{\alpha}$ for the dual quasar candidate SDSS J0222-0857 (right two panels), respectively. In each separate panel, the red and blue lines are for the two cores with similar colors as shown in Figure \ref{fig:f4}.
\label{fig:f6}}
\end{figure*}

\begin{figure*}
\plotone{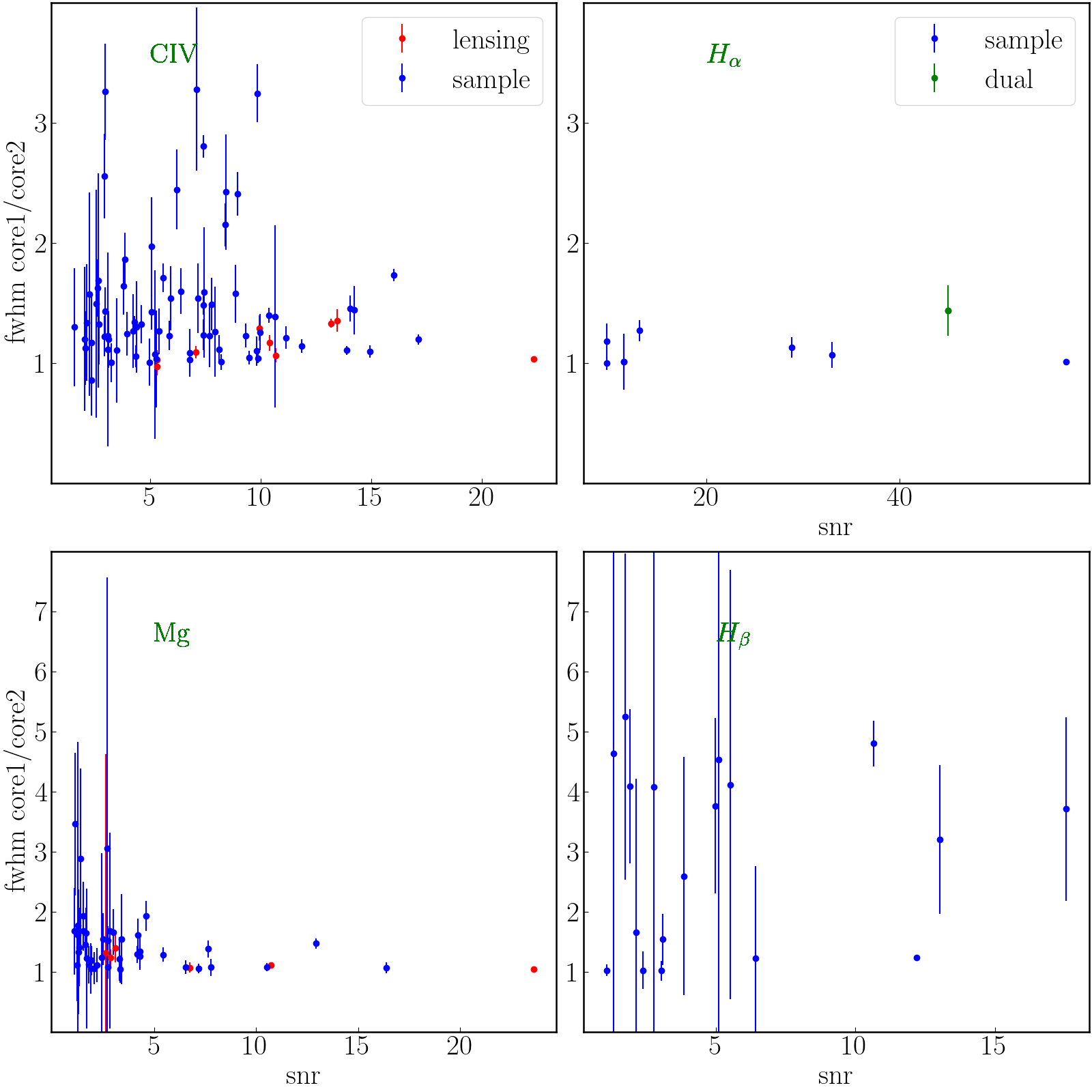}
\caption{
FWHM Comparison of the broad emission line for CIV, $H_\alpha$, MgII and $H_{\beta}$ lines versus the spectra signal to noise ratio for all the targets in our sample from the top left to the bottom right panels, respectively. In each separate panel, we mark the lensed quasars in red, the dual quasars in green and the blue colors for other targets, respectively.
\label{fig:f7}}
\end{figure*}

\subsubsection{FWHM comparison of BLs.}
\label{sec:result_fwhm}

Given the challenges in distinguishing dual quasars from lensed quasars based solely on spectra and flux ratios, particularly with varying observation times for both cores within the same system, we further compare the FWHM of the broad emission lines between the two cores. In Figure \ref{fig:f6}, we present the FWHM comparison results for the reported lensed quasar SDSS J1001 +5027 and the newly selected dual quasar candidate SDSS J0222 -0857.

For the lensed quasar SDSS J1001+5027, the FWHM ratios for the CIV and MgII lines are approximately 1.35 and 1.0, respectively. This suggests that despite the continuum flux variation between the two observation times, the FWHM of the emission lines from the two cores remains consistent. Notably, there is a distinct absorption feature near the blue end of the CIV line, probably resulting from absorption by the foreground lensing galaxy. This feature could be instrumental in exploring the characteristics of the lensing galaxy. In contrast, for the dual quasar candidate SDSS J0222-0857, the FWHM ratios for $H_{\beta}$ and $H_{\alpha}$ are approximately 2 and 1.2, respectively. Moreover, the profiles of the emission lines vary, especially for the He II line, which is nearly absent in the blue core. These variations in emission line profiles and FWHM strongly support a dual quasar scenario for this system.

Based on these fitting results, we derive the FWHM distributions for the broad emission lines CIV, MgII, $H_{\beta}$, and $H_{\alpha}$, as shown in the four panels of Figure \ref{fig:f7}. The reported lensed quasars and dual quasars are also labeled for comparison. As illustrated, lensed quasars tend to exhibit similar FWHM values for their broad emission lines. Therefore, we establish a ratio threshold of 1.5 to distinguish between dual quasar/projected quasar pairs and lensed quasars. In cases where the FWHM ratio exceeds 1.5 or there are notable differences in the broad emission line profiles between the two cores within the same system, we identify the system as either a dual quasar candidate or a projected quasar pair; otherwise, it is categorized as a lensed quasar candidate.

\subsubsection{Overview of the selection strategy}
\label{sec:result_over}

Here we present a through overview of our selection strategy:

\begin{itemize}
    \item Broad emission line profiles: Our focus is on broad emission line properties rather than continuum measurements for several reasons. First, the two cores within a system are observed at different times, which means that even in the case of lensed quasars, the continuum flux ratio can vary because of the intrinsic variability of the source. Additionally, foreground lensing galaxy absorption and microlensing effects may distort the continuum. Thus, relying on broad emission line profiles, which are less susceptible to such variability, provides a more stable feature for classification.

    \item FWHM comparison: We classify our sample by comparing the FWHM ratios between the two cores from reported lensed and dual quasars. The FWHM of the broad emission lines is derived from spectral fitting results, allowing for a quantitative comparison. However, the accuracy of this estimation is highly dependent on the quality of the observed spectra. Poor resolution or low signal-to-noise ratios can lead to errors in the FWHM measurements, potentially affecting the reliability of our classifications.

    \item Misclassifications: We also account for the possibility of misclassifications due to AGN type changes over time, known as the "changing-look AGN" scenario. In such cases, a lensed quasar may experience a shift in its AGN type, resulting in differences in broad emission line profiles between the two cores due to the time delay between observations. This could cause variations in the continuum flux ratio, emission line profiles, and FWHM between the two images, leading us to mistakenly classify the object as a dual quasar. Although changing-look AGN are rare, we cannot entirely dismiss this possibility when interpreting our results.

\end{itemize}

Our methodology offers a systematic framework for distinguishing between dual and lensed quasar candidates, albeit retaining sensitivity to data quality and the influence of rare astrophysical phenomena. Further refinement and validation, particularly via high-resolution imaging and spectra, are essential to enhance the accuracy of our classifications.

\subsection{HST coverage}

\begin{figure*}
\plotone{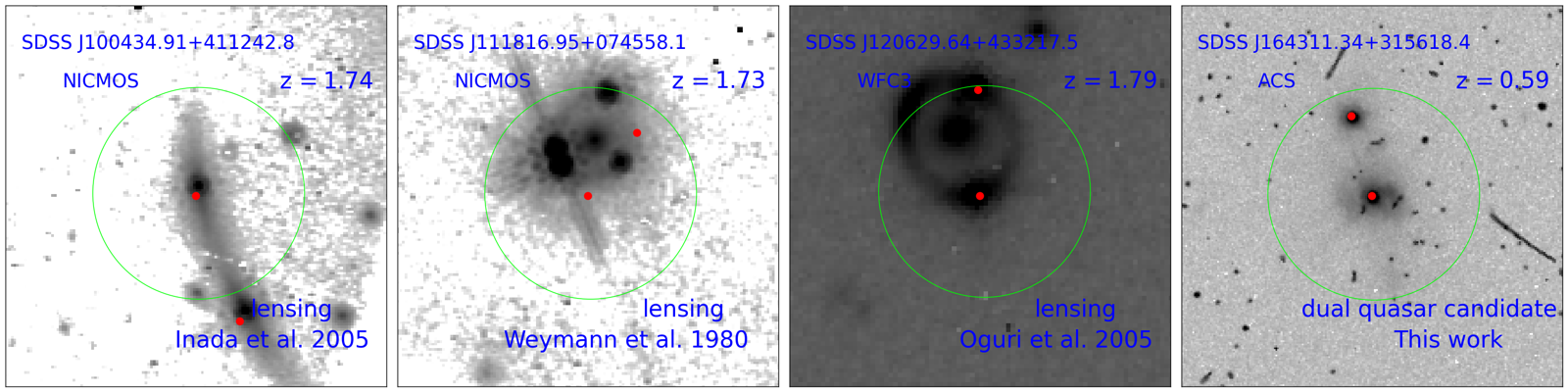}
\caption{
Archival HST images of the four targets. The resolution for each separate panel is 10 \arcsec $\times$ 10 \arcsec. In each separate panel, the green circle indicates a radius of 3 \arcsec, with the The red dots for the SDSS/DESI coordinates. The source IDs, redshift, and the classification results are labeled in the top left, top right and bottom right, respectively. We also indicate the instruments used below the source ID in each panel. 
\label{fig:f8}}
\end{figure*}

Given the high resolution of the Hubble Space Telescope (HST), it plays a crucial role in resolving the structure of both lensed quasars and dual quasars. We conducted a search of the HST archival database and identified four targets with available HST observations, as shown in Figure \ref{fig:f8}.

For the previously reported lensed quasar SDSS J111816.95+074558.1, the HST images clearly reveal multiple lensing images along with a foreground lensing galaxy at the center. The offset between the DESI coordinates and the HST coordinates might be caused by the calibration procedures for the earlier HST images. Similarly, for SDSS J120629.64+433217.5, the images show a distinct ring structure in addition to the multiple lensing images. However, in the case of SDSS J100434.91+411242.8, only two point sources resembling the lensed images are visible, and no distinct foreground lensing galaxy can be observed due to the poor quality of this target.

Apart from the three known lensed quasars, our method also uncovers a new dual quasar candidate, SDSS J164311.34+315618.4, with two distinct cores visible. No potential foreground lensing galaxy is identified in the HST ACS image, and given the intrinsic velocity offset, this is proposed as a newly identified dual quasar candidate. Overall, HST observations corroborate the precision and reliability of our classification methods.

\begin{figure*}
\plotone{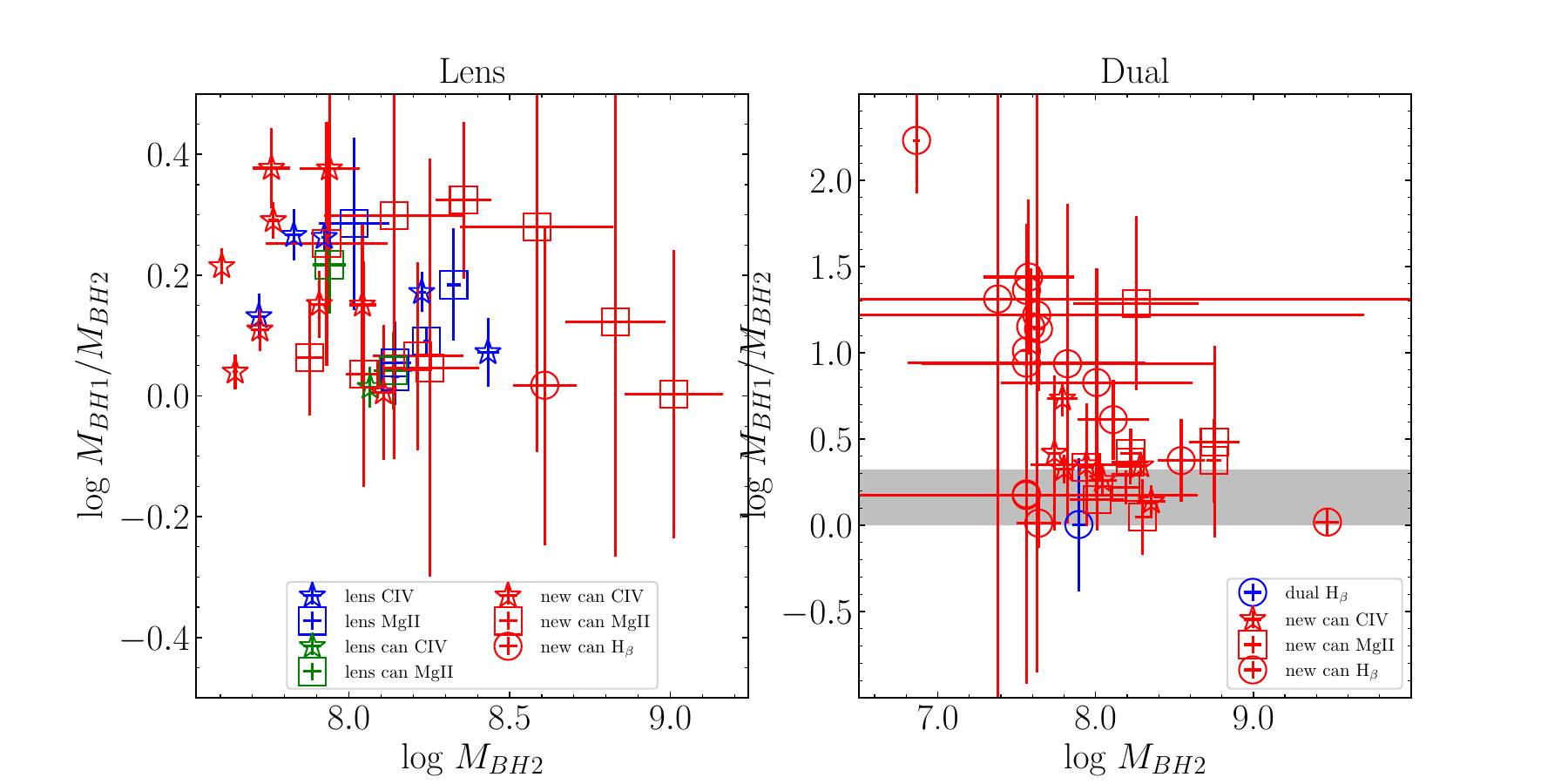}
\caption{
Black hole mass ratio between the two cores in the same system with respect to the smaller black hole's mass for selected dual and lensed quasar candidates in our sample. The left and right panels are for the lensed quasar and dual quasar scenario, respectively. In each panel, the reported lens/dual, the reported lens/dual candidates and our newly selected dual/lens candidates are presented in blue, green and red colors,  respectively. The black hole mass calculated from CIV broad emission lines, MgII broad emission lines and $H_{\beta}$ broad emission lines are shown in five-point stars, squares and circles, respectively. For reference, we also indicate the ratio scale for the lensed quasar scenario with gray regions as in the right panel for the dual quasar case.
\label{fig:f9}}
\end{figure*}

\subsection{Black hole mass ratio comparison}

We also analyze the ratio of black hole masses between the two cores in the same system for our selected dual and lensed quasar candidates, as illustrated in the left and right panels of Figure \ref{fig:f9}. 

In the case of lensed quasars (candidates), the typical range for black hole masses is from $10^{7.5}$ to $10^{9} \ M_{\odot}$, with mass ratios between the cores usually being small and within the margin of error. This evidence supports the notion that lensed quasars are deflected images of a common background source. 
Considering the underlying physics, the number of lensed quasars decreases at lower redshifts, thus making the computation of black hole masses using $H_{\beta}$ infrequent, with only one newly detected candidate at a redshift of 0.91. Our newly identified lensed quasar candidates exhibit mass distributions similar to those of previously reported lensed quasars and candidates.

In contrast, for dual quasars, black hole masses generally lie between $10^{6.5}$ and $10^{9.5} \ M_{\odot}$. The mass ratios between the cores in dual quasars show a broader range compared to lensed quasars, with discrepancies reaching exceeding 100 times for the dual quasar candidate SDSS J100508.08+ 341424.1, implying that the central black holes formed under distinct conditions. It is worth noting that while the black hole masses estimated from different broad emission lines can differ, the mass ratios are consistent across both lensed and dual quasar systems.

\section{Conclusion} 
\label{sec:conclusion}

In this study, we present a systematic approach to identifying lensed quasar and dual quasar candidates through spectroscopic analysis, applied to the SDSS DR17 and DESI EDR catalogs. We first cross-match the catalogs within a 5-arcsecond radius and visually select 200 targets that show clear evidence of two distinct cores. We process the DESI or SDSS spectra using PyQSOFit, comparing broad emission line profiles and FWHM ratios between the two cores of each system. Lensed quasars typically show consistent FWHM values between the two cores, whereas dual quasars tend to exhibit broader values. We set a threshold FWHM ratio of 1.5 to distinguish between these two types.

We identify 30 lensed quasar candidates with similar broad emission line profiles and FWHM ratios between the two cores. Among these, 8 were lensed quasars previously confirmed and 3 were reported as lensed quasar candidates in previous studies. We further find 36 dual quasar candidates, with one previously confirmed dual quasar and one dual quasar candidate reported in previous works. These dual quasars have distinct broad emission line profiles and intrinsic velocity differences of less than 600 km/s. The remaining targets are composed of projected pairs or those with low-resolution SDSS/DESI spectra.

We also explore the HST archival database and find available images for 4 targets. Three are lensed quasars previously reported. One target is a newly identified dual quasar candidate, supported by the HST image. We further investigate the black hole mass ratios between the two types of quasar pairs. For lensed quasar candidates, the black hole mass ratios between the two cores are generally consistent. However, for dual quasars, the mass ratios can vary significantly, with some cases showing differences more than 100 times. This stark contrast in mass ratio distributions underscores the fundamentally different physical nature of these two quasar types: lensed quasars are images of the same object, while dual quasars represent two distinct supermassive black holes in a merging system. With the upcoming DESI DR1 spectroscopic survey, which will provide high-resolution spectra for billions of targets, we aim to expand our research to discover more lensed and dual quasars using similar methods.

\begin{acknowledgments}
This work is supported by National Key R\&D Program of China No.2022YFF0503402. ZYZ acknowledges support by the China-Chile Joint Research Fund (CCJRF No. 1906). We also acknowledge the science research grants from the China Manned Space Project, especially, NO. CMS-CSST-2021-A04, CMS-CSST-2021-A07. 
\end{acknowledgments}

\vspace{5mm}
\facilities{SDSS, DESI, HST}

\software{PyQSOFit}

\begin{table*}[]
\centering
\footnotesize
\caption{Basic properties of the selected lensed quasar candidates. The reported lensed quasars and the reported lensed quasar candidates are marked with * and + in the upper right corner of their source ids, respectively.}
\label{table:tb2}
\begin{tabular}{llllllll} \hline \hline
name                     & ra1              & dec1              & ra2              & dec2              & $Z_{1}$   & $Z_{2}$   & sep \\ \hline
$\rm SDSS \ J121646.05+352941.5^{*}$ & 184.19188        & 35.494863         & 184.19135        & 35.494881         & 2.017 & 2.006 & 1.555      \\
$\rm SDSS \ J125418.94+223536.5^{*}$ & 193.57896        & 22.593497         & 193.57939        & 22.593322         & 3.649 & 3.65  & 1.562      \\
$\rm SDSS \ J111816.95+074558.1^{*}$  & 169.57063        & 7.7661636         & 169.57024        & 7.7666605         & 1.733 & 1.732 & 2.266      \\
$\rm SDSS \ J133907.13+131039.6^{*}$  & 204.77974        & 13.177676         & 204.78014        & 13.177413         & 2.239 & 2.236 & 1.692      \\
$\rm SDSS \ J140012.77+313454.1^{*}$  & 210.05322        & 31.581705         & 210.05355        & 31.581319         & 3.314 & 3.316 & 1.719      \\
$\rm SDSS \ J100128.61+502756.8^{*}$  & 150.36921        & 50.465805         & 150.36813        & 50.466224         & 1.841 & 1.846 & 2.898      \\
$\rm SDSS \ J120629.64+433217.5^{*}$  & 181.62353        & 43.538219         & 181.62355        & 43.539053         & 1.791 & 1.794 & 3.003      \\
$\rm SDSS \ J100434.91+411242.8^{*}$  & 151.14549        & 41.211895         & 151.14503        & 41.210908         & 1.738 & 1.731 & 3.765      \\
$\rm SDSS \ J081254.82+334950.2^{+}$ & 123.22843        & 33.830613         & 123.22845        & 33.831122         & 1.500   & 1.500   & 1.833      \\
$\rm SDSS \ J074013.44+292648.3^{+}$ & 115.05602        & 29.446782         & 115.05594        & 29.446046         & 0.978 & 0.977 & 2.661      \\
$\rm SDSS \ J221208.11+314418.8^{+}$ & 333.03380         & 31.738564         & 333.03356        & 31.737847         & 1.709 & 1.714 & 2.684      \\
SDSS J084821.79+232732.0 & 132.09082        & 23.458912         & 132.09129        & 23.459008         & 1.603 & 1.606 & 1.590       \\
SDSS J133321.77+393626.0 & 203.34012        & 39.606977         & 203.34073        & 39.607225         & 2.204 & 2.202 & 1.913      \\
SDSS J142536.29+531248.3 & 216.40123        & 53.213421         & 216.40029        & 53.213484         & 0.911 & 0.914 & 2.039      \\
SDSS J015820.41+251210.4 & 29.585059        & 25.202894         & 29.584235        & 25.202747         & 1.843 & 1.844 & 2.736      \\
SDSS J121417.42+523152.0 & 183.57258        & 52.531207         & 183.57328        & 52.531874         & 2.308 & 2.302 & 2.849      \\
SDSS J124025.15+432916.5 & 190.10479        & 43.487925         & 190.10390         & 43.487351         & 3.264 & 3.268 & 3.110       \\
SDSS J130326.17+510047.1 & 195.85903        & 51.013190          & 195.85891        & 51.014169         & 1.683 & 1.685 & 3.535      \\
SDSS J022542.41-051452.4 & 36.426717        & -5.247905        & 36.426938        & -5.246824        & 1.258 & 1.259 & 3.974      \\
SDSS J015342.79+205453.2 & 28.428309        & 20.914791         & 28.429423        & 20.915245         & 2.339 & 2.345 & 4.087      \\
SDSS J153038.56+530403.9 & 232.66068        & 53.067813         & 232.66175        & 53.066851         & 1.535 & 1.535 & 4.165      \\
SDSS J224204.63+055830.4 & 340.51930         & 5.9751224         & 340.51822        & 5.9746274         & 2.511 & 2.517 & 4.258      \\
SDSS J162902.59+372430.8 & 247.26081        & 37.408568         & 247.26098        & 37.409760          & 0.922 & 0.926 & 4.319      \\
SDSS J232624.65+331600.2 & 351.60273        & 33.266739         & 351.60129        & 33.266757         & 1.690  & 1.689 & 4.335      \\
SDSS J094004.26+522344.5 & 145.01777        & 52.395697         & 145.01575        & 52.395121         & 1.783 & 1.796 & 4.898      \\
SDSS J142402.22+343915.4 & 216.00930 & 34.654269  & 216.00960 & 34.653750  & 2.012 & 2.013 & 2.070       \\
J122811.440 +583632.04   & 187.04767 & 58.608899   & 187.04690 & 58.608262  & 1.651 & 1.653 & 2.709      \\
J1410 6.189  -01002.53   & 212.52579  & -1.00070 & 212.52506  & -1.00039 & 2.153 & 2.153 & 2.857      \\
J135554.892 +053241.99   & 208.97872 & 5.54500  & 208.97814 & 5.54589  & 1.121 & 1.126 & 3.814      \\
J100148.638 +012954.76   & 150.45266 & 1.49854  & 150.45137 & 1.49889  & 1.488 & 1.485 & 4.792  \\
\hline  \hline
\end{tabular}
\end{table*}

\begin{table*}[]
\centering
\caption{Basic properties of the selected dual quasar candidates. The reported dual quasar and the reported dual quasar candidate are marked with * and + in the upper right corner of their source ids, respectively.}
\label{table:tb3}
\footnotesize
\begin{tabular}{llllllll} \hline \hline
source id        & ra1              & dec1             & ra2              & dec2             & $Z_{1}$   & $Z_{2}$   & Sep \\ \hline
$ \rm SDSS \ J171322.58+325627.9^{*}$ & 258.34412        & 32.94112        & 258.34552        & 32.94123         & 0.102 & 0.101 & 4.25       \\
$\rm SDSS J121405.12+010205.1^{+}$ & 183.52129 & 1.03533 & 183.52138 & 1.03473 & 0.494 & 0.492 & 2.179      \\
SDSS J145307.06+331950.5 & 223.27943        & 33.330706        & 223.27953        & 33.330253        & 1.191 & 1.189 & 1.658      \\
2MASX J12033688+2006304  & 180.90382        & 20.108507        & 180.90362        & 20.108058        & 0.212 & 0.212 & 1.752      \\
SDSS J011812.03-010442.5 & 19.55013         & -1.078479        & 19.550538        & -1.0782041       & 0.74  & 0.74  & 1.771      \\
2MASS J08082144+0558351  & 122.08933        & 5.9764731        & 122.08983        & 5.97636          & 0.341 & 0.341 & 1.836      \\
SDSS J022226.11-085701.3 & 35.608844        & -8.9503658       & 35.609289        & -8.9507644       & 0.167 & 0.166 & 2.136      \\
SDSS J120303.23+222351.7 & 180.76347        & 22.397703        & 180.76411        & 22.39756         & 0.616 & 0.618 & 2.192      \\
SDSS J100508.08+341424.1 & 151.28362        & 34.240037        & 151.28288        & 34.240054        & 0.162 & 0.162 & 2.203      \\
SDSS J164311.34+315618.4 & 250.79726        & 31.938444        & 250.79745        & 31.939072        & 0.587 & 0.586 & 2.334      \\
SDSS J132130.74+535529.4 & 200.37815        & 53.924904        & 200.37712        & 53.924614        & 1.653 & 1.657 & 2.42       \\
SDSS J103204.96+523748.1 & 158.02069        & 52.63004         & 158.02154        & 52.630524        & 1.632 & 1.635 & 2.547      \\
SDSS J124257.32+254303.0 & 190.73885        & 25.717513        & 190.738          & 25.717436        & 0.826 & 0.827 & 2.771      \\
SDSS J090303.78+551032.6 & 135.76566        & 55.17583         & 135.76452        & 55.175413        & 0.361 & 0.361 & 2.783      \\
SDSS J123401.31+063214.9 & 188.50550         & 6.5374859        & 188.5052         & 6.5367127        & 0.559 & 0.559 & 2.983      \\
SDSS J142232.77+523940.1 & 215.63658        & 52.661166        & 215.63553        & 52.660602        & 1.493 & 1.494 & 3.063      \\
SDSS J081252.45+402348.8 & 123.21857        & 40.396903        & 123.21756        & 40.39648         & 0.189 & 0.188 & 3.16       \\
SDSS J114719.33+003348.4 & 176.83008        & 0.5642204       & 176.83058        & 0.56348529       & 0.263 & 0.262 & 3.201      \\
SDSS J223709.89-010248.1 & 339.29128        & -1.046662        & 339.29208        & -1.0471525       & 1.758 & 1.762 & 3.378      \\
SDSS J160633.43+515011.7 & 241.63934        & 51.836664        & 241.64088        & 51.836823        & 2.248 & 2.248 & 3.473      \\
% SDSS J094852.08+342432.5 & 147.21701        & 34.409046        & 147.21791        & 34.409761        & 2.258 & 2.267 & 3.711      \\
SDSS J091117.79+440622.2 & 137.82414        & 44.106169        & 137.82284        & 44.105665        & 0.954 & 0.958 & 3.819      \\
SDSS J100233.90+353127.5 & 150.64128        & 35.524327        & 150.64255        & 35.524634        & 2.305 & 2.304 & 3.882      \\
SDSS J093446.32+302239.9 & 143.69230         & 30.376884        & 143.69304        & 30.377754        & 0.983 & 0.985 & 3.885      \\
SDSS J012228.85+311146.8 & 20.62025         & 31.196351        & 20.618979        & 31.196239        & 0.749 & 0.751 & 3.935      \\
SDSS J223839.70+270622.0 & 339.66542        & 27.10613         & 339.66484        & 27.107182        & 1.293 & 1.293 & 4.219      \\
SDSS J105609.80+551604.1 & 164.04086        & 55.267831        & 164.0395         & 55.266897        & 0.256 & 0.257 & 4.369      \\
SDSS J235235.55+041916.2 & 358.14814        & 4.3211833        & 358.14861        & 4.3200431        & 0.700   & 0.698 & 4.438      \\
SDSS J024512.12-011313.9 & 41.300529        & -1.2205771       & 41.299757        & -1.2215581       & 2.465 & 2.459 & 4.494      \\
% SDSS J144415.35+344829.8 & 221.06397        & 34.808284        & 221.06246        & 34.808441        & 2.282 & 2.271 & 4.499      \\
SDSS J103724.07+481955.5 & 159.35031        & 48.332108        & 159.35066        & 48.330824        & 0.848 & 0.847 & 4.698      \\
SDSS J213626.51+010213.7 & 324.11048        & 1.0371713        & 324.11036        & 1.038505         & 2.06  & 2.06  & 4.821      \\
SDSS J162753.07+463724.1 & 246.97113        & 46.623369        & 246.97272        & 46.622576        & 1.968 & 1.968 & 4.858      \\
SDSS J123659.63+624956.3 & 189.24856 & 62.83223 & 189.248895 & 62.83266 & 1.496 & 1.500   & 1.643      \\
J175939.365 +625614.94   & 269.91402 & 62.93748 & 269.91328 & 62.93782 & 1.02  & 1.019 & 1.715      \\
J140221.099 +042624.93   & 210.58791 & 4.44026 & 210.58753 & 4.43966 & 1.128 & 1.128 & 2.544      \\
SDSS J123007.31+611532.9 & 187.52872 & 61.25887  & 187.53044   & 61.25913 & 0.438 & 0.437 & 3.125      \\
J140236.244 +053429.24   & 210.65102 & 5.57479 & 210.65085 & 5.57354 & 1.260  & 1.260  & 4.547     \\
\hline \hline
\end{tabular}
\end{table*}

% \begin{acknowledgments}
% To be revised later
% \end{acknowledgments}

\clearpage
\bibliography{reference}{}
\bibliographystyle{aasjournal}

\end{document}